\documentclass[utf8]{frontiersSCNSarXiv} % for Science, Engineering and Humanities and Social Sciences articles
\usepackage{url,hyperref,microtype,subcaption}
\usepackage[onehalfspacing]{setspace}

\newcommand{\hc}{{h_\mathrm{c}}}
\newcommand{\rc}{{r_\mathrm{c}}}
\newcommand{\gc}{{g_\mathrm{c}}}
\newcommand{\drm}{{\mathrm{d}}}
\newcommand{\gE}{{g_\mathrm{E}}}

\usepackage{amsmath}

\usepackage[english]{babel}

\usepackage{tikz}
\usetikzlibrary{decorations.pathreplacing}
\usetikzlibrary{positioning,arrows,shapes.geometric}
\usetikzlibrary{patterns}
\usepackage{tcolorbox}
\tcbuselibrary{skins}
\def\keyFont{\fontsize{8}{11}\helveticabold }
\def\firstAuthorLast{Maindl {et~al.}} %use et al only if is more than 1 author
\def\Authors{T.~I.\ Maindl\,$^{*}$, R.\ Miksch and B.\ Loibnegger}
\def\Address{Department of Astrophysics, University of Vienna, Vienna, Austria}
\def\corrAuthor{Department of Astrophysics, University of Vienna, T\"urkenschanzstr.\ 17, 1180 Vienna, Austria}

\def\corrEmail{thomas.maindl@univie.ac.at}

\begin{document}
\onecolumn
\firstpage{1}

\title[Stability of a rotating asteroid housing a space station]{Stability of a rotating asteroid housing a space station} 

\author[\firstAuthorLast ]{\Authors} %This field will be automatically populated
\address{} %This field will be automatically populated
\correspondance{} %This field will be automatically populated

\extraAuth{}% If there are more than 1 corresponding author, comment this line and uncomment the next one.
%\extraAuth{corresponding Author2 \\ Laboratory X2, Institute X2, Department X2, Organization X2, Street X2, City X2 , State XX2 (only USA, Canada and Australia), Zip Code2, X2 Country X2, email2@uni2.edu}

\maketitle

\begin{abstract}
%
%%% Leave the Abstract empty if your article does not require one, please see the Summary Table for full details.
\section{}
%For full guidelines regarding your manuscript please refer to \href{http://www.frontiersin.org/about/AuthorGuidelines}{Author Guidelines}.
%
%As a primary goal, the abstract should render the general significance and conceptual advance of the work clearly accessible to a broad readership. References should not be cited in the abstract. Leave the Abstract empty if your article does not require one, please see \href{http://www.frontiersin.org/about/AuthorGuidelines#SummaryTable}{Summary Table} for details according to article type. 
%
Today there are numerous studies on asteroid mining. They elaborate on selecting the right objects, prospecting missions, potential asteroid redirection, and the mining process itself. For economic reasons, most studies focus on mining candidates in the 100--500\,m size-range. Also, suggestions regarding the design and implementation of space stations or even colonies inside the caverns of mined asteroids exist. Caverns provide the advantages of confined material in near-zero gravity during mining and later the hull will shield the inside from radiation. Existing studies focus on creating the necessary artificial gravity by rotating structures that are built inside the asteroid. Here, we assume the entire mined asteroid to rotate at a sufficient rate for artificial gravity and investigate its use for housing a habitat inside. In this study we present how to estimate the necessary spin rate assuming a cylindrical space station inside a mined asteroid and discuss the implications arising from substantial material stress given the required rotation rate. We estimate the required material strength using two relatively simple analytical models and apply them to fictitious, yet realistic rocky near-Earth asteroids.

\tiny
 \keyFont{ \section{Keywords:} asteroids, asteroid mining, material stress, arificial gravity, space stations} %All article types: you may provide up to 8 keywords; at least 5 are mandatory.
\end{abstract}

\section{Introduction}\label{sec:intro}

%For Original Research Articles \citep{conference}, Clinical Trial Articles \citep{article}, and Technology Reports \citep{patent}, the introduction should be succinct, with no subheadings \citep{book}. For Case Reports the Introduction should include symptoms at presentation \citep{chapter}, physical exams and lab results \citep{dataset}.

Sustaining human life on a station built inside a mined asteroid is a task which will require expertise in many fields. There needs to be air to breathe, water to drink, and the appropriate recycling systems, as well as food and light. Nevertheless, one of the most important prerequisites for a human body to stay healthy is gravity. 

Taking plants to space and zero gravity is not as big a problem as taking a human body into this hostile environment. Plants do adapt to zero gravity relatively easily which is shown by many experiments performed both on the ISS (International Space Station) as well as on the ground with artificial microgravity  \citep[for example, see][]{kitaya00,kitaya01,kiss09}.

%Calcium from the bones secretes out through urine which weakens the bones like people with osteoporosis

The human body however, reacts sensitive to zero gravity. The lack of an up and down direction in space affects endothelial cells \citep{versari13}, blood distribution to endocrine and reflex mechanisms controlling body water homeostasis and blood pressure \citep[e.g.][]{taibbi13}, and it results in muscle wasting \citep{golpa10}, severe bone loss \citep{keyak09}, immune depression \citep[e.g.][]{cogoli93,battista12,gasperi14}, and ophthalmic problems \citep{nelson14}.

%Investigations on medical issues for humans on a mission to Mars and on the Martian surface have been undergone by \citeasnoun{jones04} and \citeasnoun{korienek98}. 
A study on how much gravity is needed to keep the human body upright was performed by \citet{harris14}. They found that the threshold level of gravity needed to influence a persons orientation judgment is about 15\,\% of the gravity on Earth's surface, which is approximately the gravity acting on the Lunar surface. Martian gravity, 38\,\% of Earth's gravity, should be enough for astronauts to orient themselves and maintain balance. 

As a consequence of a lack of experiments on the influence of reduced gravity on the human body we adopt the value of 38\,\% of Earth's gravity ($\gE$) as starting point for our theoretical approach. We assume that a rotation of the asteroid has to cause an artificial gravity of minimum $0.38 \, \gE$  in order to sustain long term healthy conditions for humans on the station. 

Present suggestions to tackle this challenge of providing sufficient gravity rely on habitats in rotating wheels or tori that create gravity: \citet{grabaz13} suggest self-sustained colonies up to 2000 people. Other studies somewhat vaguely mention \emph{augmenting the natural rotation with additional artificial rotation\/} \cite{taygra08}. We elaborate on the latter and explore the feasibility and viability of creating “artificial” gravity for a habitat by putting the entire asteroid to rotation at a rate sufficient to generate the desired gravity.

We start with initial considerations regarding the required spin rate of a space station with sufficient artificial gravity (Sect.~\ref{sec:initialcons}). Section~\ref{sec:stress} elaborates on the stress acting on the asteroidal hull and formulates two analytical models for the tensile and shear stresses as a function of the asteroid size, the dimensions of the space station, the required artificial gravity level, and the bulk density of the asteroid material. Section~\ref{sec:usecase} applies our formulation to a near-Earth asteroid maximizing the usable area of the space station while observing maximum stress constraints. Finally, conclusions and an outlook on further related research is given in Sect.~\ref{sec:conclusions}.

\section{Initial considerations}\label{sec:initialcons}

\begin{figure}[h!]
\centering
\includegraphics[width=0.25\linewidth]{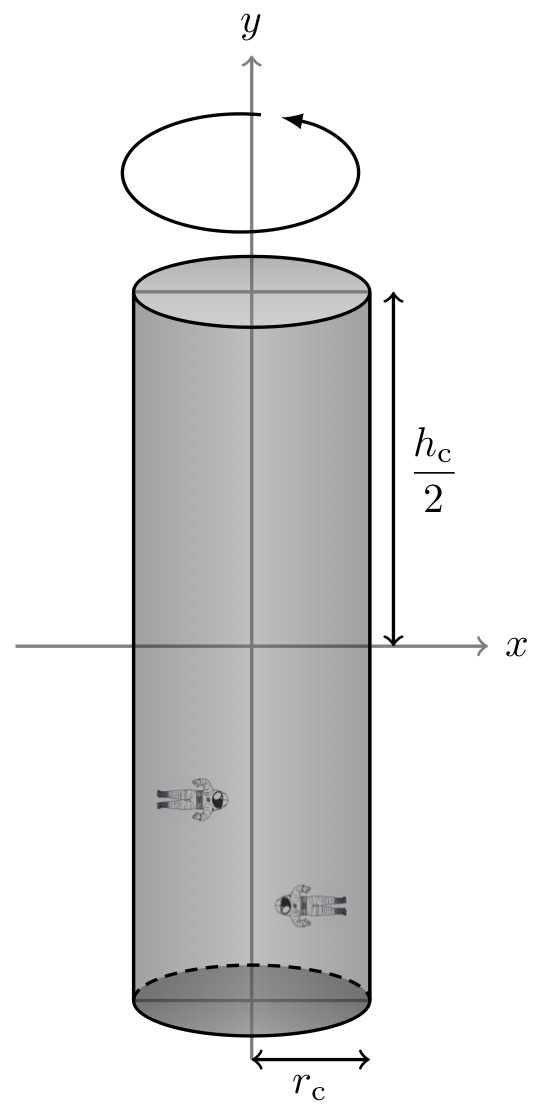}
\caption{Sketch of a cylindrical space station of height $\hc$ and radius $\rc$.} \label{fig:cyl}
\end{figure}

\begin{figure}[h!]
    \centering
    \includegraphics[width=0.75\linewidth]{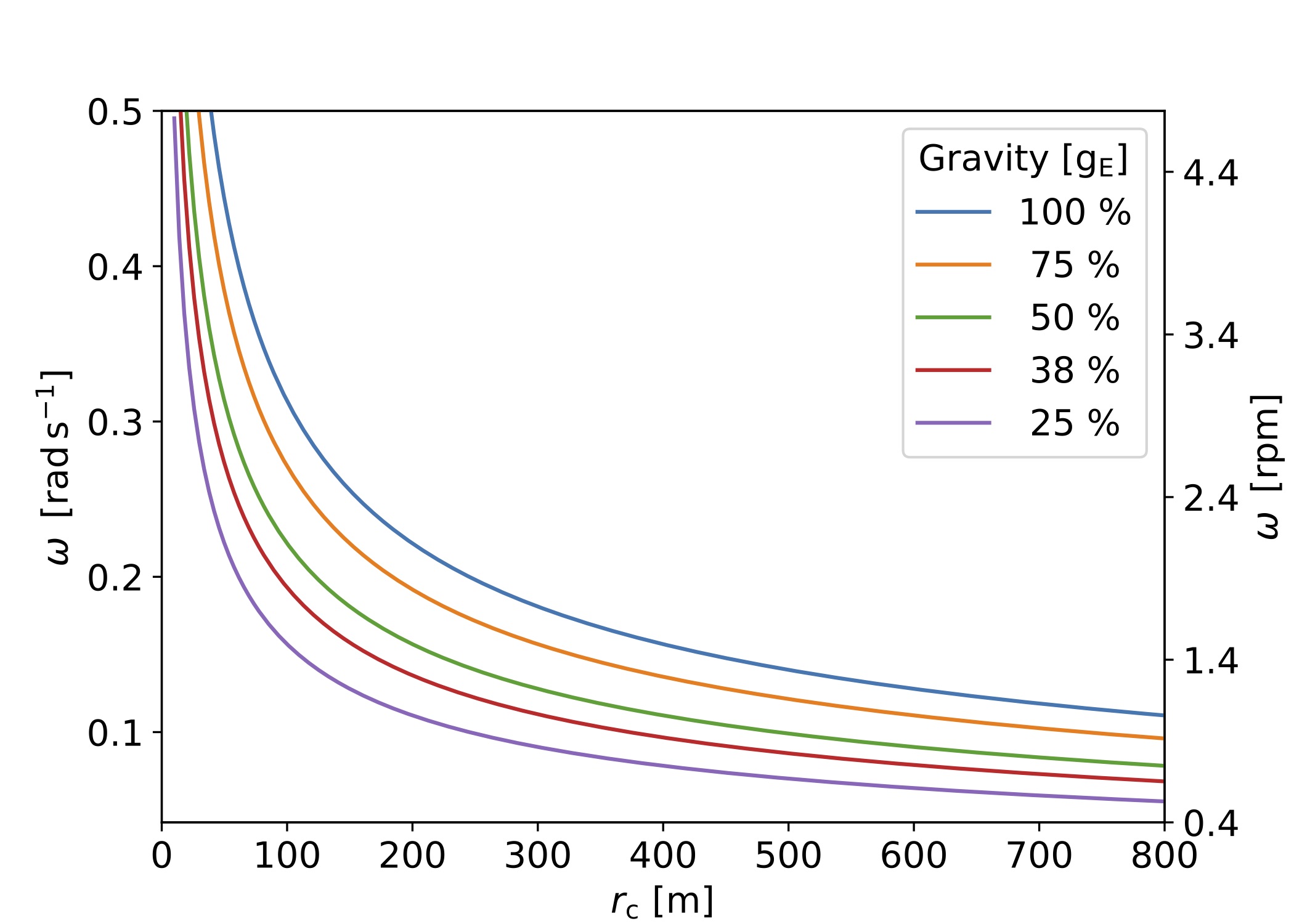}
    \caption{Rotation rates $\omega\/$ in rad\,s$^{-1}$
    and rpm necessary to achieve an artificial gravity of $\gc$ on the lateral surface of a cylinder with radius $\rc$.}
    \label{fig:omegaT}
\end{figure}

Let's assume a cylindrical space station with height $\hc$ and radius $\rc$ as depicted in Fig.~\ref{fig:cyl}. Letting the cylinder rotate about its symmetry axis $y\/$ with angular velocity $\omega\/$ will create an acceleration of
\begin{equation}
\gc=\omega^2\rc
\end{equation}
acting on objects on the lateral surface. For a certain artificial gravity level the required rotation rate $\omega\/$ and rotation period $T\/$ are then given by
\begin{equation}
\omega = \sqrt{\frac{\gc}{\rc}},\qquad T = \frac{2\pi}{\omega} = 2\pi\sqrt{\frac{\rc}{\gc}}.\label{eq:omega}
\end{equation}
Considering a size range of $\rc=50\ldots 250\,\mathrm{m}$ for example, the rotation rates would need to be between 1.17 and 2.6\,rpm (rotations per minute) to create artificial gravity necessary for sustaining extended stays on the station (assuming $0.38 \, \gE$ as discussed above). Figure~\ref{fig:omegaT} gives an overview of rotation rates for a range of radii and gravity levels. The area usable for a space station subject to artificial gravity is the lateral surface of the cylinder $S\/$,
\begin{equation}
    S=2\pi\,\rc\,\hc.\label{eq:S}
\end{equation}

\section{Estimating tensile and shear stress}\label{sec:stress}

The station is to be built inside a mined asteroid. Imposing a spin rate sufficient for providing artificial gravity on the lateral surface of the cylinder will create a substantial load on the asteroid material due to centrifugal forces. While little is known on material properties of small asteroids subject to our study, we rely on assumed material strength. Here, we assume that the asteroid is made of homogeneous, solid material such as basaltic silicate rock, for instance. We estimate the load on the asteroid material in simplified models: the tensile stress acting on the asteroid cross section is related to assumed tensile strength of solid silicate rock.

Figure~\ref{fig:1} shows our geometrical model of a spheroidal asteroid with semi-axes $a\/$ and $b\/$, respectively. The cavern is centered and cylindrical with respective radius $\rc$ and height $\hc$. Due to the elliptical cross-section, the lateral distance $d\/$ between the cavern and the asteroid's surface is given by
\begin{equation}
    d=\frac{b}{a}\,\sqrt{a^2-\frac{\hc^2}{4}}-\rc.
\end{equation}
For a feasible solution, the cavern needs to be inside the asteroid in its entirety, which translates to the condition $d>0$ or
\begin{equation}
    \rc < \frac{b}{a}\,\sqrt{a^2-\frac{\hc^2}{4}}.\label{eq:inside}
\end{equation}
The whole body will rotate about its symmetry axis $y\/$ at a rate $\omega\/$ providing sufficient artificial gravity on the lateral surface of the space station with radius $\rc$. We assume rigid body rotation.

\subsection{Model 1}\label{sec:model1}

\begin{figure}[h!]
\centering
\includegraphics[width=0.5\linewidth]{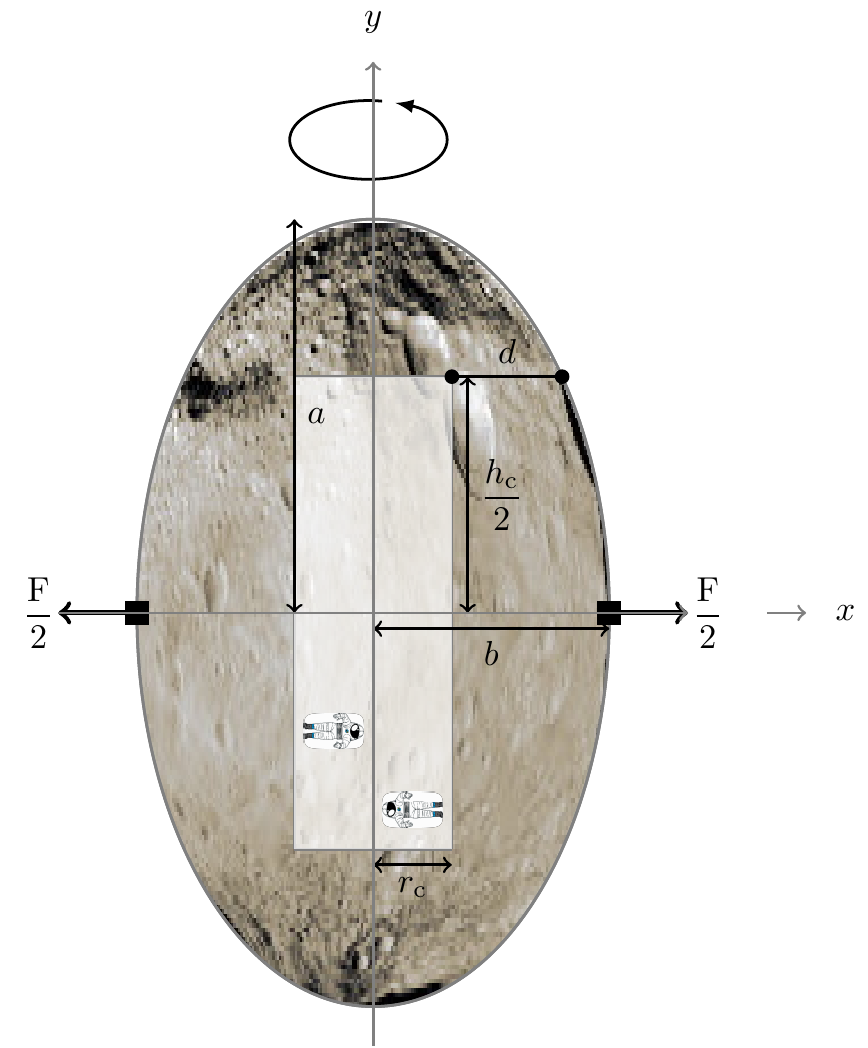}
\caption{Sketch of asteroid with cylindrical cavern used to house a space station. Picture: Asteroid Vesta. Credit: NASA/JPL-Caltech/UCAL/MPS/DLR/IDA.} \label{fig:1}
\end{figure}

\begin{figure}[h!]
    \centering
    {\includegraphics[width=0.9\linewidth]{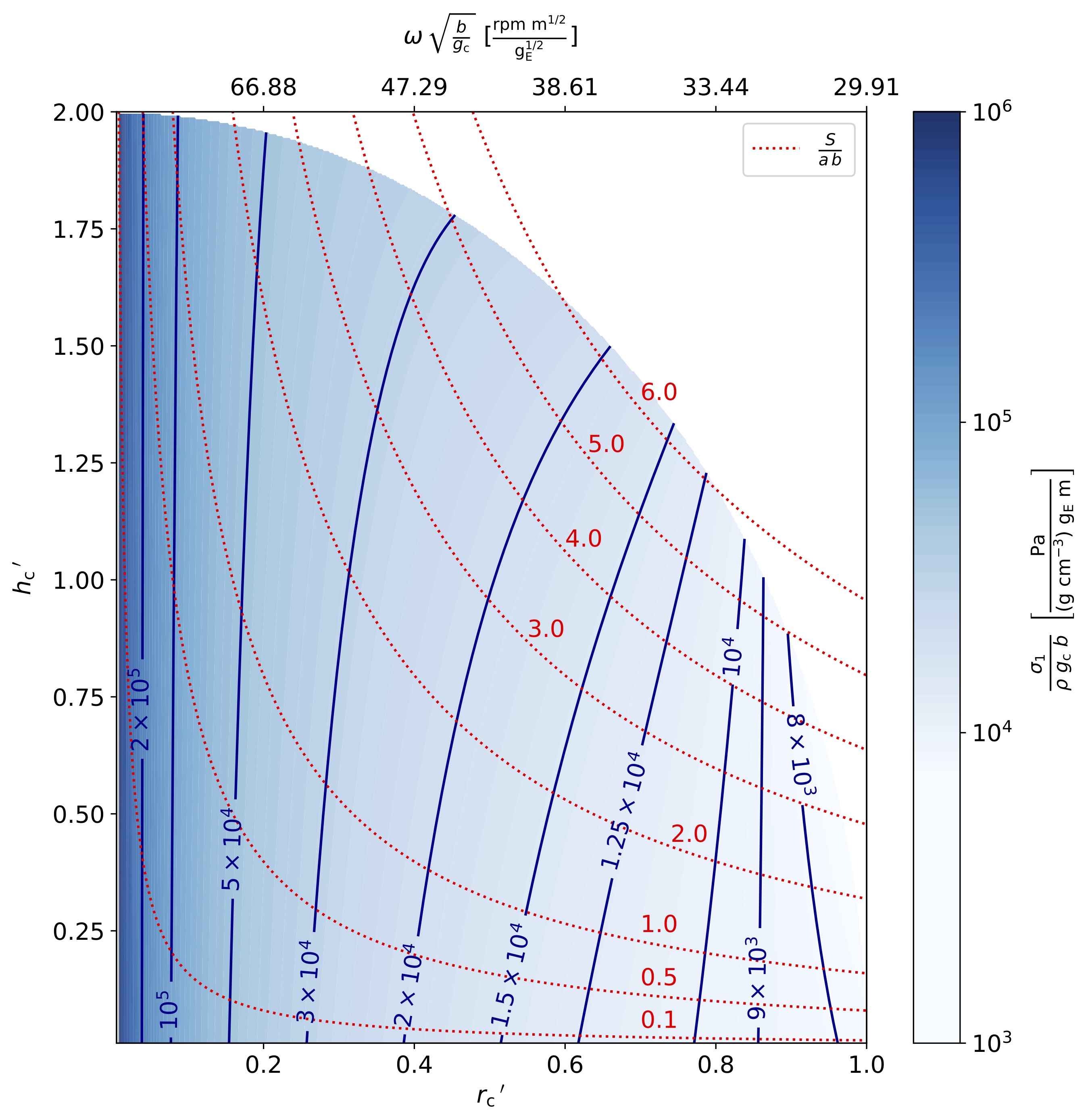}}
    \caption{Model 1 material stress for different dimensions of the space station (dimensionless radius $\rc$ and height $\hc$). The blue contour lines give the ratio $\frac{\sigma}{\rho\,\gc b}\/$ for $\rho\/$ given in $\mathrm{g\,cm^{-3}}$, $\gc$ measured in units of $\mathrm{g_E}$, and $b\/$ measured in meters, respectively. The dotted red lines give contours of the ratio $\frac{S}{a\,b}\/$ obtained via (\ref{eq:Sprime}), the upper $x\/$-axis gives the scaled rotation rate.}
    \label{fig:generic}
\end{figure}

In this model centrifugal forces that are acting on the asteroid material exert a load on an arbitrary symmetry plane. We determine the tensile stress that results from this load. The total centrifugal force $F_1\/$ pulling two halves of the asteroid apart is given by
\begin{equation}
    F_1=2\!\!\!\int\limits_\mathrm{asteroid\atop half}\!\!\! \mathrm{d}F=2\!\!\!\int\limits_\mathrm{asteroid\atop half}\!\!\! \mathrm{d}(m\omega^2r)=2\omega^2\rho \!\!\!\int\limits_\mathrm{asteroid\atop half}\!\!\! r\,\mathrm{d}V.
\end{equation}
Here, we use the mass $m\/$ of a volume element, the uniform asteroid density $\rho\/$, and the distance $r\/$ of a volume element from the rotation axis. Transforming into cylindrical coordinates symmetrical w.r.t.\ the $y\/$-axis ($\drm V=r\,\drm r\,\drm y\,\drm\varphi$) and considering that there will be no contribution from the void of height $\hc$ and radius $\rc$ (see Fig.~\ref{fig:1}) gives
\begin{equation}
    F_1=2\omega^2\rho\left[\int\limits_0^{\pi}\drm \varphi \int\limits_{-a}^a\drm y \int\limits_0^{\frac{b}{a}\sqrt{a^2-y^2}}r^2\drm r - \int\limits_0^{\pi}\drm \varphi \int\limits_{-\hc/2}^{\hc/2}\drm y \int\limits_0^{\rc}r^2\drm r\right].\label{eq:Fstart}
\end{equation}
Note that $r(y)=\frac{b}{a}\sqrt{a^2-y^2}$ holds for the elliptical cross section. Integrating (details given in Appendix~\ref{sec:apcent}) yields
\begin{equation}
    F_1=\pi\,\omega^2\rho\left(\frac{\pi}{4}a\,b^3-\frac{2}{3}\rc^3\hc\right).\label{eq:F}
\end{equation}
This load acts on the asteroid's cross section $A=\pi\,a\,b-2\,\rc\,\hc$ (cf.\ Fig.~\ref{fig:1}) resulting in tensile stress $\sigma_1=F_1/A$ of
\begin{eqnarray}
\sigma_1 = \frac{F_1}{A} = \frac{\pi\,\omega^2\rho}{12}\left( \frac{3\pi a\,b^3-8\rc^3\hc}{\pi\,a\,b-2\,\rc\,\hc}\right).\label{eq:sigma1}
\end{eqnarray}
As we are interested in the stress resulting from a desired artificial gravity $\gc$ we substitute $\omega^2$ from (\ref{eq:omega}) and get
\begin{eqnarray}
\sigma_1 = \frac{\pi\,\rho\,\gc}{12\rc}\left( \frac{3\pi a\,b^3-8\rc^3\hc}{\pi\,a\,b-2\,\rc\,\hc}\right).\label{eq:sigma1g}
\end{eqnarray}
By introducing the dimensionless quantities
\begin{equation}
    \hc'=\frac{\hc}{a}, \qquad \rc'=\frac{\rc}{b}
\label{eq:hprimerprime}
\end{equation}
we can separate the parameters specific to individual asteroids and show that $\sigma\/$ scales linearly with each of material density $\rho\/$, desired artificial gravity $\gc\/$, and asteroid semi-minor axis $b\/$. Inserting $\hc'$ and $\rc'$ into (\ref{eq:sigma1g}) yields
\begin{equation}
    \sigma_1 = \frac{\pi\,\rho\,\gc\,b}{12\,\rc'}\left(\frac{3\pi -8\,\rc'^{\,3}\hc'}{\pi-2\,\rc'\,\hc'}\right)
    = \rho\,\gc\,b\cdot f(\rc',\hc').\label{eq:sigma1prime}
\end{equation}
Hence, it is sufficient to study $f(\rc',\hc')\/$ to get estimates for the stresses in asteroids of arbitrary density and semi-minor axis rotating at a rate providing the desired artificial gravity. Figure~\ref{fig:generic} shows $\sigma_1$ contours assuming parameters $\rho=1\,\mathrm{g\,cm^{-3}}$, $\gc=1\,\mathrm{g_E}$, and $b=1\,\mathrm{m}$. For different parameter values the numbers scale linearly with $\rho$, $\gc$, and $b$. The white area in Fig.~\ref{fig:generic} corresponds to illegal combinations of $\rc$ and $\hc$ that violate condition (\ref{eq:inside}) demanding the space station has to be inside the asteroid in its entirety.

The required rotation rate as a function of space station radius $\rc$ and desired artificial gravity $\gc$ is given in (\ref{eq:omega}), scales according to
\begin{equation}
    \omega=\sqrt{\frac{\gc}{\rc}}=\sqrt{\frac{\mathrm{g_E}}{b}}\sqrt{\frac{\gc[\mathrm{g_E}]}{\rc'}},
\end{equation}
and is indicated on the upper $x\/$-axis. As we will be interested in the usable surface area of the station $S\/$, we indicate its dimensionless variant
\begin{equation}
    \frac{S}{a\,b}=2\pi\,\rc'\hc'\label{eq:Sprime}
\end{equation}
as red, dotted contour lines in Fig.~\ref{fig:generic}. For any assumed maximum material strength, the solution for maximum $S\/$ suggests a radius of the cylinder $\rc$ that extends all the way to the surface ($d=0$, cf.\ Fig.~\ref{fig:1}). This is however, the edge case of our model 1 that estimates material load by assuming two halves of the asteroid driven apart by centrifugal forces. As soon as $d\/$ gets very small the ``two halves'' assumption fails. Therefore, we estimate material load by another model which will be more accurate at the edge case, i.e.\ very small values of the distance to the surface $d\/$.

\subsection{Model 2}\label{sec:model2}

\begin{figure}[h!]
\centering
\includegraphics[width=0.35\linewidth]{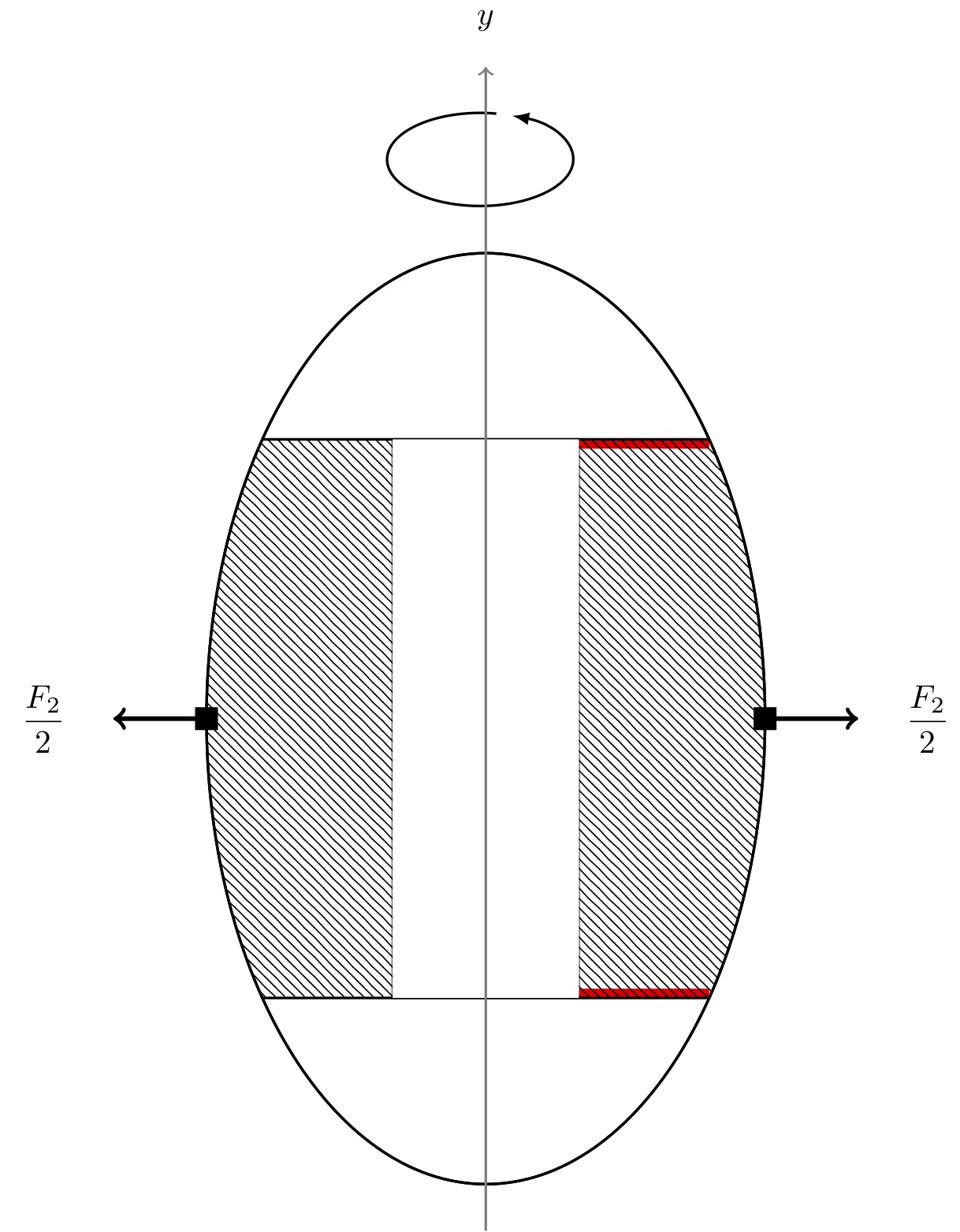}
\caption{Simplified sketch of cavity for clarification. Force acting on material.} \label{fig:dm}
\end{figure}

Rather than focusing on two entire halves of the hollowed-out asteroid, this model studies the ``mantle'' outside the space station. This is the solid torus created by sweeping the right part of the hashed surface between the red lines at $y=\pm\,\hc/2$ in Fig.~\ref{fig:dm} around the $y\/$-axis. As the radius of the space station approaches the asteroid's hull, the centrifugal forces attempting to shear away this torus may get significant. This shearing load acts on the two annuli resulting from rotating the red lines in Fig.~\ref{fig:dm} about the $y\/$-axis. In addition to overcoming the shear strength of the asteroid material however, the tensile strength of the cross section (the hashed area in Fig.~\ref{fig:dm}) has to be exceeded by the load exerted by the centrifugal force.

Similar to calculating $F_1\/$ in Sect.~\ref{sec:model1}, the centrifugal force $F_2\/$ can be derived by transforming to cylindrical coordinates as follows:

\begin{align}
    F_2&=\omega^2\rho \!\!\!\!\!\int\limits_{\mathrm{shaded\ volume}\atop{\mathrm{in\ Fig.~\ref{fig:dm},}\atop \mathrm{i.e.\ }|y|\le\hc/2}}\!\!\!\!\! r\,\mathrm{d}V
    =\omega^{2} \rho\!\! \int\limits_{|y|\le \frac{1}{2}\hc}\!\! r^{2} \ \mathrm{d}r \ \mathrm{d}y \ \mathrm{d}\varphi %\\
%    &
= \omega^{2} \rho \int\limits_{0}^{2\pi} \mathrm{d}\varphi \ \int\limits_{-\frac{1}{2}\hc}^{\frac{1}{2}\hc} \mathrm{d}y \int\limits_{\rc}^{\frac{b}{a} \sqrt{a^2 - y^2}}\!\! \mathrm{d}r \ r^2.\label{eq:FMinitial}
\end{align}
Integrating (Appendix~\ref{sec:apFM} gives the detailed steps) yields a rather lengthy expression for the centrifugal force:
%\begin{eqnarray}
%\fl F_M= \frac{2\pi}{3} \omega^2 \rho\,a\,b^3 \left\{\rule{0mm}{8mm}\right.\frac{1}{8} \left[\rule{0mm}{8mm}\right. \hc' \left(5 - \frac{\hc'\,^2}{2} \right) \sqrt{1 - \frac{\hc'\,^2}{4}} + 6 \arcsin \frac{\hc'}{2} \left.\rule{0mm}{8mm}\right] - \hc' \ \rc'\,^3 \left.\rule{0mm}{8mm}\right\}.
%\end{eqnarray}

\begin{eqnarray}
F_2 = \frac{2\pi\omega^2\rho}{3} \left\{\rule{0mm}{8mm}\right.\frac{1}{8} \left( \frac{b}{a}\right)^3 \left[\rule{0mm}{8mm}\right. \hc \left(5a^2 - \frac{\hc^2}{2} \right) \sqrt{a^2 - \frac{\hc^2}{4}} + 6 a^4 \arcsin \frac{\hc}{2a} \left.\rule{0mm}{8mm}\right] - \hc \ \rc^3 \left.\rule{0mm}{8mm}\right\}.\label{eq:FM}
\end{eqnarray}
This force exerts a tensile load on the asteroid's cross section $A^t$ between $y=-\hc/2$ and $y=\hc/2$ and a shear load on the two annuli of area $A^s$ given by rotating the red lines in Fig.~\ref{fig:dm} about the $y\/$-axis. The surface area $A^t$ is given by (cf.\ Fig.~\ref{fig:dm})

\begin{align}
A^t&=2\left[\int\limits_{-\hc/2}^{\hc/2}x(y)\,\mathrm{d} y-\hc\rc\right] = 4\int\limits_{0}^{\hc/2}\frac{b}{a}\sqrt{a^2-\frac{y^2}{4}}\,\mathrm{d} y-2\hc\rc\nonumber\\
&=\frac{4b}{a}\left[\frac{y}{4}\sqrt{4a^2-y^2}+a^2\arctan\frac{y}{\sqrt{4a^2-y^2}}\right]_0^\frac{h}{2} -2\hc\rc\nonumber
\end{align}
and using the identity $\arcsin x = \arctan (x/\sqrt{1-x^2)}$ we get

\begin{eqnarray}
A^t=\frac{b\,\hc}{2a}\sqrt{4a^2-\frac{\hc^2}{4}}+4ab\arcsin\frac{\hc}{4a}-2\hc\rc.\label{eq:AT}
\end{eqnarray}
Each of the annuli has a surface of $A^s$,

\begin{align}
    A^s&=\pi\left[(\rc+d)^2-\rc^2\right]=\pi\left[\frac{b^2}{a^2}\,\left(a^2-\frac{\hc^2}{4}\right)-\rc^2\right]\\
    &=\pi\left[b^2\left(1-\frac{\hc^2}{4a^2}\right)-\rc^2\right] \label{eq:AS}%\\
%    A^s=\pi\,b^2\left(1-\frac{\hc'\,^2}{4}-\rc'\,^2\right)
.
\end{align}
Combining equations (\ref{eq:FM}), (\ref{eq:AT}), and (\ref{eq:AS}), we obtain the average stress $\sigma_2$ in this model,

\begin{eqnarray}
\sigma_2=\frac{F_2}{A^t+2A^s},
\end{eqnarray}
for the complete -- rather unwieldy -- formulation of the stress please refer to Appendix~\ref{sec:apsigma2}. Unlike it is the case in model 1, we cannot formulate $\sigma_2$ in a scaling way using the dimensionless quantities $\rc$ and $\hc$. For the asteroids in scope of this study though, $\sigma_2$ is usually smaller than $\sigma_1$, but increases if the space station radius gets closer to the asteroid's surface. In the following Sect.~\ref{sec:usecase} we will demonstrate this by comparing the two models for a fictitious, yet realistic asteroid.

\section{Application to a realistic asteroid}\label{sec:usecase}

We will apply the analytic models 1 and 2 to a rocky asteroid with dimensions $500\times 390\,\mathrm{m}$. There is a number of similar-sized rocky near-Earth asteroids, e.g. 3757 Anagolay, 99942 Apophis, 3361 Orpheus, 308635 (2005 YU55), 419624 (SO16), etc.\ \citep[cf.][]{JPL18}. As little is known about the composition and material properties of these objects, we assume they are composed of basaltic rock with a bulk density of $\rho=2.7\,\mathrm{g\,cm^{-3}}$. Tensile strength values for basalt are in the range of approx.\ 12\ldots 14\,MPa \cite{sto69}, shear strengths are approx.\ 8\ldots36\,MPa \cite{karcih15}, which provides an order-of-magnitude framework of the expected material strength data. Finally, we will assume a desired artificial gravity level of $\gc=0.38\,\mathrm{g_E}$ as discussed in Sect.~\ref{sec:intro}.

\begin{figure}[h!]
    \centering
    \includegraphics[width=0.75\linewidth]{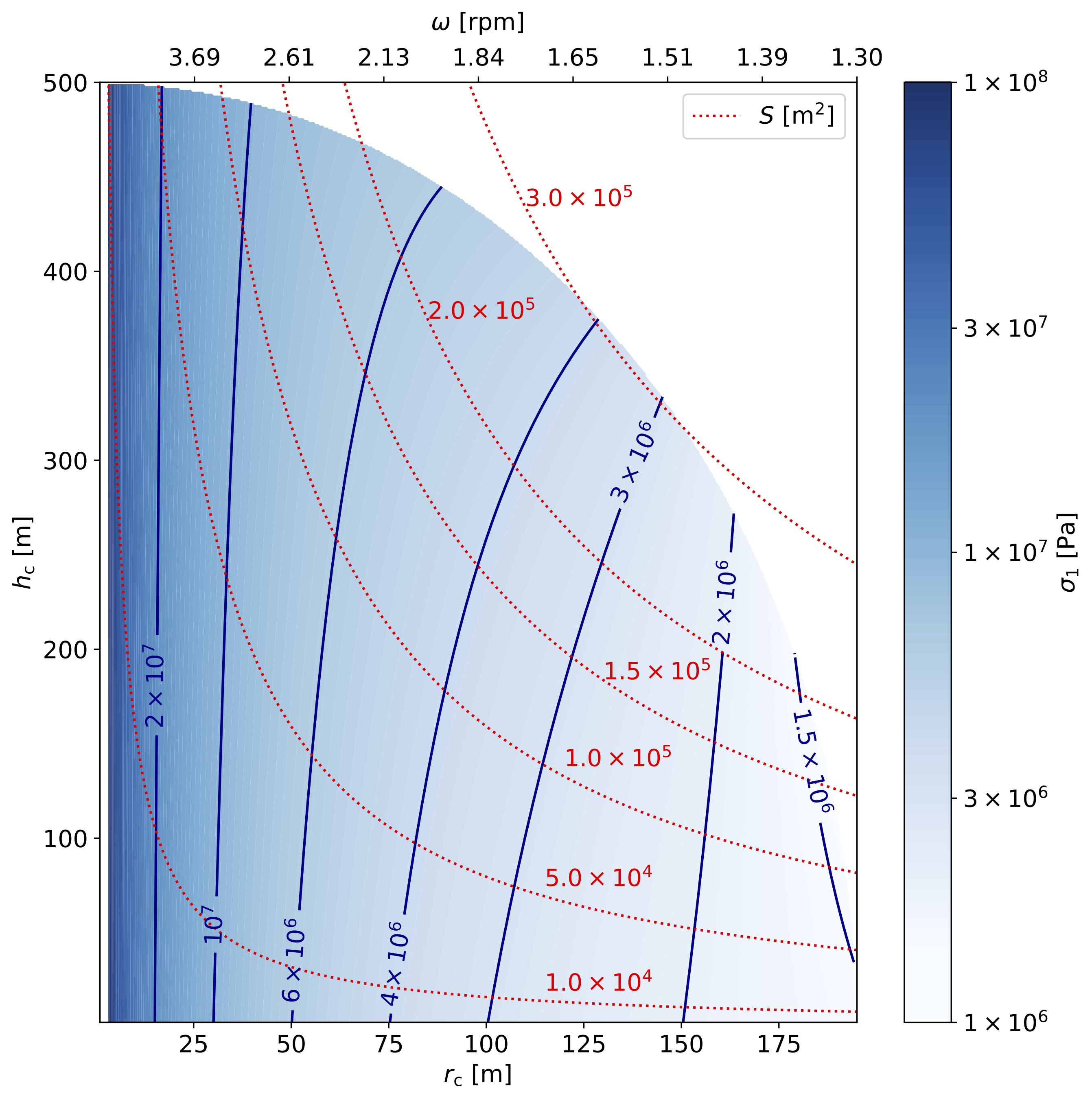}
    \caption{Model 1 results for artificial gravity of 0.38\,$\mathrm{g_E}$ in a space station of radius $\rc$ and height $\hc$, the color code and the blue contour lines give the tensile stress $\sigma_1$ resulting from the required rotation rate obtained via (\ref{eq:sigma1}). The red dotted lines give the usable surface area $S\/$ of the space station, the upper $x\/$-axis denotes the required rotation rate.}
    \label{fig:anagolaytensile}
\end{figure}

Figure~\ref{fig:anagolaytensile} shows the resulting tensile stress along with the usable space station surface and required rotation rates predicted by model 1. The stress levels are mostly of the same order of magnitude as the assumed material strength ($\sim 10$\,MPa) or even smaller. However, the solution resulting in the maximum area $S\approx 0.3\,\mathrm{km^2}$ would have the cylindrical station extend to the asteroid's surface, which seems unrealistic and will lead to the asteroid becoming unstable. Also, for realistic scenarios a material stress ($\approx 4$\,MPa in this case) very close to the -- poorly constrained -- material strength will be unacceptable so that a cavern with radius-height data more towards the lower right of the diagram will be desirable.

\begin{figure}[h!]
    \centering
    \includegraphics[width=0.75\linewidth]{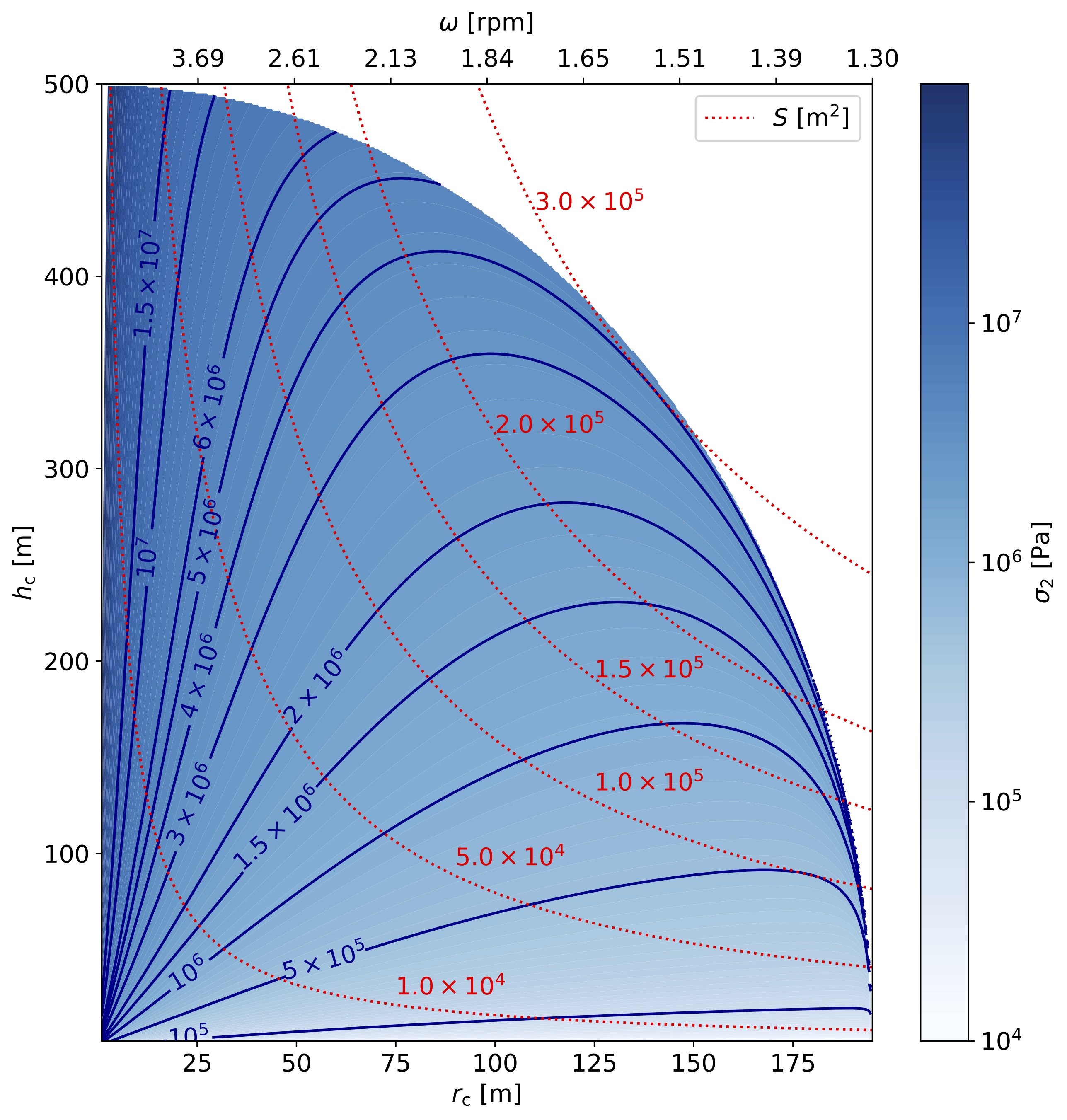}
    \caption{Model 2 results for artificial gravity of 0.38\,$\mathrm{g_E}$ in a space station of radius $\rc$ and height $\hc$, the color code and the blue contour lines give the combined tensile and shear stress resulting from the required rotation rate obtained via (\ref{eq:sigma1}). The red dotted lines give the usable surface area $S\/$ of the space station, the upper $x\/$-axis denotes the required rotation rate.}
    \label{fig:anagolaymodel2}
\end{figure}

Investigating the combined tensile and shear stresses according to our model 2 results in a different stress-pattern, given in Fig.~\ref{fig:anagolaymodel2}. While the material loads are systematically lower for space stations deeper inside the asteroid in their entirety, stresses for ``thinner'' tori (i.e., larger $\rc$) are of the same order of magnitude as predicted by model~1.

In summary, both models predict stresses that are comparable to anticipated material strength for asteroids made of competent rock. As the assumed material parameters are based on the unknown composition, thorough studies of candidate asteroids will be necessary before considering to set them to rotation to house a space station with artificial gravity inside. 

\section{Conclusions and further research}\label{sec:conclusions}

We established two simple analytical models for estimating whether a candidate for asteroid mining may be suitable for hosting a space station with artificial gravity. The novelty in our approach is to investigate whether the asteroidal hull -- once set to rotation as a whole -- can sustain the material loads resulting from a sufficiently high rotation rate.  We find that loads resulting from centrifugal forces are in the order of magnitude of material strength of solid rock, which makes a space station in the cavern of a mined asteroid feasible if its dimensions are chosen right and if the material composition and material strength of the asteroid is known to a satisfactory level of accuracy. Practical applications will crucially depend on knowing not only the composition but also the internal structure of candidate bodies. As missions to these asteroids seem inevitable for such studies, decisions on inhabiting such asteroids may only be possible after mining operations have started. Also, the methods of actually initiating the rotation at the required rate is subject to further investigations. Hypothetically, starts and landings of spacecraft during the mining process might contribute to building up angular momentum of the asteroid.   

Currently, we are working on a more realistic analytic approach for determining the detailed shape of the cavern housing the space station taking into account the internal density profile.

In the past, we successfully conducted smooth particle hydrodynamics (SPH) simulations of asteroids \citep[e.g.][]{maisch13,hagmai17,maiegg19}. As our analytical study is approximative in nature we plan to conduct a series of SPH simulations with different material models and varying porosity. This will allow to numerically verify the predictions of the simplified analytical models presented here as well as future models and to further investigate the behavior of rotating bodies with substantial internal caverns.

\section*{Conflict of Interest Statement}
%All financial, commercial or other relationships that might be perceived by the academic community as representing a potential conflict of interest must be disclosed. If no such relationship exists, authors will be asked to confirm the following statement: 

The authors declare that the research was conducted in the absence of any commercial or financial relationships that could be construed as a potential conflict of interest.

\section*{Author Contributions}

TIM developed the analytical models and performed most of the calculations. He wrote about 70\,\% of the paper and created most of the figures. RM provided various aspects of the analytical approximations, helped in the calculations, and contributed to the representation of the individual equations. He wrote about 15\,\% of the paper. BL contributed to the quality of the analytical models, researched the details of required artificial gravity levels, and created the pictorials of the asteroid with the cylindrical cavern. She wrote about 15\,\% of the paper.

%The Author Contributions section is mandatory for all articles, including articles by sole authors. If an appropriate statement is not provided on submission, a standard one will be inserted during the production process. The Author Contributions statement must describe the contributions of individual authors referred to by their initials and, in doing so, all authors agree to be accountable for the content of the work. Please see  \href{http://home.frontiersin.org/about/author-guidelines#AuthorandContributors}{here} for full authorship criteria.

\section*{Funding}
This project received seed funding from the Dubai Future Foundation through Guaana.com open research platform.
The authors also acknowledge support by the FWF Austrian Science Fund project S11603-N16.

\section*{Acknowledgments}
% This is a short text to acknowledge the contributions of specific colleagues, institutions, or agencies that aided the efforts of the authors.

The authors are indebted to Dr.\ C.\ M.\ Sch\"afer for carefully checking the formulae and wish to thank Drs.\ \'A.\ Bazs\'o and C.\ Lhotka for many fruitful discussions and constructive feedback.

\appendix
\section*{Appendix}
\renewcommand{\thesection}{\Alph{section}}
\section{Model 1 centrifugal force calculation}\label{sec:apcent}
Starting at (\ref{eq:Fstart}),

\begin{eqnarray}
F_1=2\omega^2\rho\left[\int\limits_0^{\pi}\drm \varphi \int\limits_{-a}^a\drm y \int\limits_0^{\frac{b}{a}\sqrt{a^2-y^2}}r^2\drm r - \int\limits_0^{\pi}\drm \varphi \int\limits_{-\hc/2}^{\hc/2}\drm y \int\limits_0^{\rc}r^2\drm r\right]
\end{eqnarray}
we get

\begin{align}
F_1 & = \frac{2\pi}{3} \omega^{2} \rho \left[\left(\frac{b}{a}\right)^3\int\limits_{-a}^{a} \drm y  \left(a^2 - y^2\right)^\frac{3}{2} - \rc^3 \int\limits_{-\frac{1}{2}\hc}^{\frac{1}{2}\hc} \drm y\right] \\
& = \frac{2\pi}{3} \omega^{2} \rho\left\{\left(\frac{b}{a}\right)^3 \frac{1}{8} \left[\rule{0mm}{8mm} y(5a^2-2y^2) \sqrt{a^2-y^2} + 3a^4 \arctan \frac{y}{\sqrt{a^2 - y^2}} \right]_{-a}^a - \rc^3\, \hc \right\}.
\end{align}
Considering

\begin{equation}
    \lim_{x\to\pm\infty}\arctan x = \pm \frac{\pi}{2},
\end{equation}
this simplifies to

\begin{eqnarray}
F_1 = \frac{2\pi}{3} \omega^2 \rho \left[ \left(\frac{b}{a}\right)^3 \frac{3}{8}\,\pi a^4 - \rc^3\, \hc \right] = \pi\,\omega^2\rho\left(\frac{\pi}{4}a\,b^3-\frac{2}{3}\rc^3\hc\right).
\end{eqnarray}
This is the same expression as (\ref{eq:F}).

\section{Model 2 centrifugal force calculation}\label{sec:apFM}
Starting at (\ref{eq:FMinitial}),

\begin{eqnarray}
F_2 = \omega^{2} \rho \int_{0}^{2\pi} \mathrm{d}\varphi \ \int_{-\frac{1}{2}\hc}^{\frac{1}{2}\hc} \mathrm{d}y \ \int_{\rc}^{\frac{b}{a} \sqrt{a^2 - y^2}} \mathrm{d}r \ r^2
\end{eqnarray}
we get

\begin{align}
F_2 & = 2\pi \omega^{2} \rho \int_{-\frac{1}{2}\hc}^{\frac{1}{2}\hc} \drm y \int_{\rc}^{\frac{b}{a} \sqrt{a^2 - y^2}} \drm r \ r^2 \\
& = 2\pi \omega^{2} \rho \int_{-\frac{1}{2}\hc}^{\frac{1}{2}\hc} \drm y \ \frac{1}{3} \left[ \left( \frac{b}{a}\right)^3 \left(a    ^2 - y^2 \right)^{\frac{3}{2}} - \rc^3 \right] \\
&= \frac{2\pi}{3} \omega^2 \rho \left[ \left( \frac{b}{a}\right)^3 \int_{-\frac{1}{2}\hc}^{\frac{1}{2}\hc} \drm y \left(a^2 - y^2 \right)^{\frac{3}{2}} - \hc\, \rc^3 \right] \\
&= \frac{2\pi}{3} \omega^2 \rho \left\{\rule{0mm}{8mm}\right. \left( \frac{b}{a}\right)^3 \frac{1}{8} \left[\rule{0mm}{8mm}\right. y \left(5a^2 - 2y^2 \right) \sqrt{a^2 - y^2} %\nonumber\\
 + 3a^4 \arctan \left(\frac{y}{\sqrt{a^2 - y^2}}\right) \left.\rule{0mm}{8mm}\right]_{-\frac{1}{2}\hc}^{\frac{1}{2}\hc} - \hc\,\rc^3 \left.\rule{0mm}{8mm}\right\}.
\end{align}
With the identity $\arcsin x = \arctan (x/\sqrt{1-x^2)}$ and $\arcsin(-x)=-\arcsin x$ the integral evaluates to

\begin{align}
F_2&= \frac{2\pi}{3} \omega^2 \rho \left\{\rule{0mm}{8mm}\right.\frac{1}{8} \left( \frac{b}{a}\right)^3 \left[\rule{0mm}{8mm}\right. \frac{\hc}{2} \left(5a^2 - 2 \ \frac{\hc^2}{4} \right) \sqrt{a^2 - \frac{\hc^2}{4}} \nonumber\\
&\qquad\qquad + 3 a^4 \arcsin \frac{\hc}{2a} + \frac{\hc}{2} \left(5a^2 - 2 \ \frac{\hc^2}{4} \right) \sqrt{a^2 - \frac{\hc^2}{4}} - 3 a^4 \arcsin \frac{-\hc}{2a} \left.\rule{0mm}{8mm}\right] - \hc \ \rc^3 \left.\rule{0mm}{8mm}\right\}\\
&= \frac{2\pi\omega^2 \rho}{3}  \left\{\frac{1}{8} \left( \frac{b}{a}\right)^3 \left[\rule{0mm}{8mm}\right. \hc \left(5a^2 - \frac{\hc^2}{2} \right) \sqrt{a^2 - \frac{\hc^2}{4}} + 6 a^4 \arcsin \frac{\hc}{2a} \left.\rule{0mm}{8mm}\right] - \hc \ \rc^3 \right\}.
\end{align}
This is the same expression as (\ref{eq:FM}).

\section{Model 2 stress}\label{sec:apsigma2}
Combining equations (\ref{eq:FM}), (\ref{eq:AT}), and (\ref{eq:AS}), we get the total stress $\sigma_2$,

\begin{align}
\sigma_2&=\frac{F_2}{A^t+A^s}\\
&=\frac{2\pi\omega^2\rho}{3} \left\{\frac{1}{8} \left( \frac{b}{a}\right)^3 \left[ \hc \left(5a^2 - \frac{\hc^2}{2} \right) \sqrt{a^2 - \frac{\hc^2}{4}} + 6 a^4 \arcsin \frac{\hc}{2a} \right] - \hc \ \rc^3 \right\}\nonumber\\
&\quad\times\left\{\frac{b\,\hc}{2a}\sqrt{4a^2-\frac{\hc^2}{4}}+4ab\arcsin\frac{\hc}{4a}-2\hc\rc+
2\pi\left[b^2\left(1-\frac{\hc^2}{4a^2}\right)-\rc^2\right]\right\}^{-1}.
\end{align}

% \section*{Supplemental Data}
%  \href{http://home.frontiersin.org/about/author-guidelines#SupplementaryMaterial}{Supplementary Material} should be uploaded separately on submission, if there are Supplementary Figures, please include the caption in the same file as the figure. LaTeX Supplementary Material templates can be found in the Frontiers LaTeX folder.

% \section*{Data Availability Statement}
% The datasets [GENERATED/ANALYZED] for this study can be found in the [NAME OF REPOSITORY] [LINK].
% % Please see the availability of data guidelines for more information, at https://www.frontiersin.org/about/author-guidelines#AvailabilityofData

\bibliographystyle{frontiersinSCNS_ENG_HUMS} % for Science, Engineering and Humanities and Social Sciences articles, for Humanities and Social Sciences articles please include page numbers in the in-text citations
\bibliography{mbrbib}

%%% Make sure to upload the bib file along with the tex file and PDF
%%% Please see the test.bib file for some examples of references

\section*{Figure captions}

%%% Please be aware that for original research articles we only permit a combined number of 15 figures and tables, one figure with multiple subfigures will count as only one figure.
%%% Use this if adding the figures directly in the mansucript, if so, please remember to also upload the files when submitting your article
%%% There is no need for adding the file termination, as long as you indicate where the file is saved. In the examples below the files (logo1.eps and logos.eps) are in the Frontiers LaTeX folder
%%% If using *.tif files convert them to .jpg or .png
%%%  NB logo1.eps is required in the path in order to correctly compile front page header %%%

% \begin{figure}[h!]
% \begin{center}
% \includegraphics[width=10cm]{logo1}% This is a *.eps file
% \end{center}
% \caption{ Enter the caption for your figure here.  Repeat as  necessary for each of your figures}\label{fig:1}
% \end{figure}

% \begin{figure}[h!]
% \begin{center}
% \includegraphics[width=15cm]{logos}
% \end{center}
% \caption{This is a figure with sub figures, \textbf{(A)} is one logo, \textbf{(B)} is a different logo.}\label{fig:2}
% \end{figure}

%%% If you are submitting a figure with subfigures please combine these into one image file with part labels integrated.
%%% If you don't add the figures in the LaTeX files, please upload them when submitting the article.
%%% Frontiers will add the figures at the end of the provisional pdf automatically
%%% The use of LaTeX coding to draw Diagrams/Figures/Structures should be avoided. They should be external callouts including graphics.

\end{document}